\documentclass[structabstract]{aa}  
\usepackage{graphicx}
\usepackage{txfonts}
\usepackage{natbib}
\begin{document}
   \title{AGN environments: is the viewing angle sufficient to explain the difference between
broad-line and narrow-line AGN?}

   \subtitle{A low-redshift study of close AGN neighbours. Paper I.}

   \author{Beatriz Villarroel \inst{1}
	  \and Andreas Korn \inst{1}\and Yoshiki Matsuoka \inst{2}
         }

  \institute{Uppsala Astronomical Observatory\\
              SE-751 20 Uppsala, Sweden\\
              \email{beatriz.villarroel@physics.uu.se, andreas.korn@physics.uu.se} \and 
	Graduate School of Science, Nagoya University, Furo-cho, Chikusa-ku, Nagoya
	464-8602, Japan\\
	\email{matsuoka@a.phys.nagoya-u.ac.jp}
       }

   \date{Received xxx 2012}

 
  \abstract
   {}
   {The unification of active galactic nuclei (AGN) is a model that has been difficult to test due to the lack of knowledge on the intrinsic luminosities of the objects. We present a test were we probe the model by statistical investigation of the neighbours to AGN at redshifts 0.03 $<$ z $<$ 0.2 within a projected distance of 350 kpc and $|\Delta$z$|<0.001$, 0.006, 0.012 and 0.03 between AGN and neighbour.}
   {1658 Type-1 (broad-line) AGN-galaxy pairs and 5698 Type-2 AGN-galaxy pairs with spectroscopic redshifts from the Data Release 7 of Sloan Digital Sky Survey were used together with a complementary set of pairs with photometric redshifts on the neighbour galaxies (13519 Type-1 AGN-galaxy and 58743 Type-2 AGN-galaxy pairs). Morphologies for the AGN host galaxies were derived from the Galaxy Zoo project.}
   {Our results suggest that broad-line AGN and narrow-line AGN reside in widely different environments where the neighbours to Type-2 AGN are more star-forming and bluer than those of Type-1 AGN. There is a colour-dependency only detectable in the neighbours with photometric redshifts for the Type-2 AGN. We see that the ratio between Type-1/Type-2 neighbours to Type-2 AGN decreases steadily at short separations with a statistical significance of 4.5 sigma. The lack of change in the morphology of the Type-2 AGN hosts having a close companion (contrary to the case of Type-1 AGN hosts) suggests that the innate state of Type-2 AGN is extremely short-lived and is not preserved in subsequent mergers. Finally, we perform a hypothetical luminosity test to investigate whether a mass bias in our selection could explain the observed differences in our samples. Our conclusion is that AGN unification is consistently not supported by the environment of the two types of AGN, but that an evolutionary connection between them might exist.}
   {}


   \maketitle
\section{Introduction}

Active galactic nuclei (AGN) are compact, very luminous objects in the center of galaxies that have an engine driven by accretion of material upon a supermassive black hole (SMBH) \citep[e.g.][]{Rees1984,LyndenBell1969}. The discovery that AGN-hosting Seyfert 2 galaxies observed with polarized light \citep{AntonucciMiller} had a similar spectrum to Seyfert 1 suggested that the two types of objects were actually the same, but that the light of a Seyfert 2 had to pass throught optically thick material on the way to us. 

The broad-line region (BLR) of the Type-1 AGN is suggested to originate from UV or X-ray illumination of hydrogen clouds around the accretion disk. The narrow-line region (NLR), consisting of narrow forbidden lines such as [\ion{O}{iii}]5007 and [\ion{N}{ii}]6585 is present in all AGN. The typical signature of Type-2 AGN is the presence of NLR and absence of BLR. This resulted in the model of AGN unification, where the accretion disk of an AGN is obscured by a donut-shaped torus of dust that from certain viewing angles hides the broad-line region around the accretion disk, making the object appear as a Type- 2 AGN \citep{Antonucci1993}. The Type-1 AGN are in this viewed face-on, where the high velocities of the clouds will cause Doppler broadening of the spectral lines. In its most simple form of the AGN Unification, the Straw Person Model (SPM), all AGN are the same type of objects only differing by the viewing angle.

While this model has been very successful in explaining some observed features of AGN such as the existence of Type-2 AGN with hidden BLRs, a number of inconsistencies also appeared with time. Unresolved issues includes e.g. why not all narrow-line AGN show broad-lines in their polarized spectra \citep{Tran2001,Tran2003}. The lack of Type-2 AGN with BLR could be explained by an extremely low accretion rate \citep{Nicastro2000,Elitzur2008}, by extreme obscuration \citep{Shu2008} or due to the relative luminosity of AGN versus the luminosity to its host galaxy making it more difficult to detect the BLR \citep{Lumsden2001}. Other difficulties are dealing with the close environment of AGN. The lack of Type-1 AGN in clusters \citep{Martinez2008} and in isolated pairs of galaxies \citep{Gonzalez2008}, the higher frequency of close companions around Type-2 AGN within 100 kpc than around Type-1 AGN \citep{Koulouridis2006a, Koulouridis2006b} and correlation of host galaxy morphology with AGN type \citep{Martinez2008} have suggested that the viewing angle solely might be insufficient to explain the difference between the two types of AGN. 

A recent study by Koulouridis et al. (2011) performing optical spectroscopy and imaging of 22 Seyfert 1 galaxies with 15 neighbours, 22 Seyfert 2 galaxies with 13 neighbours and 24 Bright IRAS galaxies (BIRGs) with 20 neighbours followed-up by X-ray spectroscopy suggested that the neighbours to Seyfert 1 and Seyfert 2 galaxies have different stellar masses, star-formation history and that Seyfert 2 neigbours have a systematically higher degree of ionization. However, their samples were too small to study any direct influence of the AGN on their neighbours.
 
Several of these findings suggest that the viewing angle might not be a sufficient explanation. This would need to be confirmed by a robust statistical study with galaxy surveys. However, when doing unification statistics one must also consider possible selection biases that a clumpy dust torus
introduces \citep{KrolikBegelman1988,Tristram2007}. The Realistic Unification \citep{Elitzur2012} takes the clumpiness of the torus into account and suggests that Type-2 AGN are more likely
to be drawn from the higher end distribution of the dust covering factor, has some observational support \citep[e.g.][]{RamosAlmeida2009,RamosAlmeida2011}. Undoubtedly, this can influence any unification statistics studies and bias us to select Type-2 AGN with larger instrinsic luminosities and SMBH masses than a Type-1 AGN sample with seemingly similar absolute magnitude distribution.

The aim of our paper is to perform the first statistical test of the AGN unification using AGN neighbours. This approach permits many biases based on observed properties of AGN to be reduced and is also the first study to combine the investigation of the direct influence of AGN on their neighbours. Hence, it offers an alternative approach to the AGN-starburst connection. In Paper I, we will use both volume-limited samples with spectroscopic redshifts drawn from the seventh data release (DR7) of the SDSS for studying the close environments of Type-1 and Type-2 AGN at a scale up to 350 kpc. We will focus on examining the clustering of neighbours around the two different types of AGN, the morphologies of the host galaxies and how colours and star-formation rates in the neighbours change as a function of their distance to their neighbour at redshifts $z <$ 0.2 using Standard Cosmology ($\Omega_{\Lambda}$=0.70, $\Omega_{M}$=0.30, H$_{o}$=70 km s$^{-1}$ Mpc$^{-1}$). From the PhotoZ catalogue, we will also obtain all neighbours with photometric redshifts and $\delta_{z} <$ 0.03 from the spectroscopically preselected AGN in order to increase our sample sizes.

Finally, we perform an extensive hypothetical experiment to deal with the potential
biases in the selection of Type-2 AGN with high obscuration and higher intrinsic luminosity and masses, and to check if the signal-to-noise-ratio of the emission lines might bias our selection and results. In Paper II (in prep.), we will include LINERs and perform a similar study using the \ion{O}{ii} lines for classification of AGN and extract the stellar masses as a replacement of the hypothetical luminosity test.

Table \ref{SumTables} summarizes all the samples in our study and their definitions. Section 2 describes our sample selection, Section 3 shows the direct results for our samples and the results from the hypothetical experiment, Section 4 is the discussion and summarizes our conclusions.

\section{Data and methods}

\subsection{Spectroscopic sample selection}

The SDSS (corrected for Galactic foreground extinction \citep{York2000} DR7 \citep{Abazajian2008} has spectroscopic data of 929,555 galaxies and 121,363 quasars in five optical bands (u,g,r,i,z filters), covering the complete wavelength range of the CCD camera. The quasars and galaxies were chosen in the redshift range 0.03 $<$ z $<$ 0.2. 
The upper redshift limit cut was selected so that H$\alpha$ could be detected in the neighbour galaxies, while the lower redshift cut was selected to avoid biases associated with finite spectroscopic fiber sizes. We used objects with specClass= 2 or 3 ('Galaxies' or 'Quasars'), $EW(H\alpha)$ $>$ 2 \AA\ and highly reliable redshifts ($zconf$ $>$0.95). We rejected those with poor quality indicated by catalog flags: brightness (flags\&0x2=0), saturation (flags\&0x40000=0) and blended images (flags\&0x8=0) \citep[see Table 9]{Stoughton02}.
The reason why we reject objects with $EW(H\alpha)$ $<$ 2 \AA\  is that our selection methods for Type-2 AGN will be based purely on emission-line criteria and we therefore must restrict our Type-1 AGN sample
to only emission-line objects for consistency. This will as a consequence do our study focused particularly
on star-forming AGN.

As we in an earlier publication \citep{Villarroel2012} observed a loss of 5\% of the quasars due to flags, we consider the bias from flagging being a negligible effect that will not significantly influence the number of AGN. 

We retrieved the redshifts, spectral line information and Galactic foreground extinction-corrected apparent magnitudes ('dered'), resulting in 253,352 objects from which we derive the parent AGN samples. For the objects marked as 'Galaxies' we retrieved K-corrections from the PhotoZ table. For the objects marked as 'Quasars' we calculated the K-correction using a universal power-law spectral energy distribution (SED) with optical flux given by $f_{\nu}=\nu^{\alpha}$ where the mean spectral optical index $\alpha$=-0.5 can be related to the K-correction using:

\begin{equation}
K(z)=-2.5 \alpha\  log(1+z) - 2.5 log(1+z)
\end{equation}\label{eq1}

The Galactic extinction- and K-corrected rest frame absolute magnitudes of all objects could then be calculated according to:

\begin{equation}
M_{abs}=m_{obs} + 5 - 5log(D_L) - K(z)     
\end{equation}\label{eq2}

\subsubsection{LINERs}

Low-ionization nuclear emission-line regions (LINERs) occur in a relatively large fraction of nearby galaxies of different morphologies and luminosities. These regions have an emission-line spectrum with low-ionization states. Emission lines from higher ionization stages are weak or absent. LINERs are subjects of scientific debate, since it unsure whether these nuclear emission regions arise from AGN activity or from star-forming regions. Neither is the mechanism behind the low ionization known, and both shockwaves and UV light are argued to be the main reason. LINERs are commonly referred to as AGN in scientific literature. However, we will separate them in our study with the help of the definition used in Kewley et al. (2006):

\begin{equation}
\log([\ion{O}{iii}]/H\beta) > 0.61/(\log([\ion{N}{ii}]/H\alpha))-0.47)+1.19 \label{eq,LINER}
\end{equation}\label{LINER}

Using the Kewley criterion we obtain 34249 LINERs in our parent sample.
Here we will only use them to compare the redshift distributions of the three types of AGN. The selection criterion will mainly be used in this paper for removing the LINERs from the Type-1 and Type-2 AGN samples. We define this sample as the "LINER Parent Sample".

\subsubsection{Type-1 AGN}

Broad Balmer emission lines is the key signature of an accretion disk. The Type-1 AGN were therefore selected on solely one criteria, the width of the H$\alpha$ line in the SDSS single-Gaussian fitted spectra, measured in the emitter's rest frame.

Galaxies should fulfill $\sigma$(H$\alpha$) $>$ 10 \AA\  (corresponding to FWHM $>$ 1000 km s$^{-1}$) and emission in H$\alpha$ in order to be categorized as Type-1 AGN. LINERs were excluded by using the Kewley criterion. The first selection gave 11334 Type-1 AGN. We define this sample as the "Type-1 Parent Sample".

The SDSS single-Gaussian fitted spectra can bias the separation between Type-1 and Type-2 AGN, since it might be difficult to resolve the contribution to the H$\alpha$ emission-line peak from the \ion{N}{ii} contribution from a narrow-line AGN. Type-2 AGN can also get "mixed" into the sample in case the broad-line component is faint, if we have scattered light from a broad-line region in an obscured quasar or if the forbidden lines in a Type-2 AGN would have a non-Gaussian line profile \citep{Liu2009}. We will address the first of these three problems later in this paper by comparing neighbours depending on the S/N ratio of the H$\alpha$ line and equivalent widths of the AGN hosts. 

We also tried out another limit, $\sigma$(H$\alpha$) $>$ 15 \AA\, to select the Type-1 AGN. The same analysis results were obtained.

\subsubsection{Type-2 AGN}

The observed high-excitation state of the gas caused by non-thermal radiation can be used to separate the AGN from starburst galaxies. To select narrow-line AGN we use the Baldwin-Phillips \& Terlevich line-ratio diagrams \citep{BPT1981} 
combined with the Kauffmann et al. (2003) criterion:

\begin{equation}
\log([\ion{O}{iii}]/H\beta) > 0.61/(\log([\ion{N}{ii}]/H\alpha))-0.05)+1.3 \label{eq,BPT}
\end{equation}\label{BPTequation}

To avoid contamination of Type-1 AGN in the Parent Sample, we only include those that also have $\sigma$(H$\alpha$) $<$ 10 \AA .  These include many composite objects also hosting starburst activity.  LINERs were excluded by the use of the Kewley criterion. The first selection yielded 53416 Type-2 AGN, defined as the "Type-2 Parent Sample". We will also for the Type-2 AGN sample check the impact of the S/N ratio of the emission line on the conclusions.

\subsection{Volume-limited test sample of parent objects}

\begin{figure*}
 \centering
  \includegraphics[scale=.7]{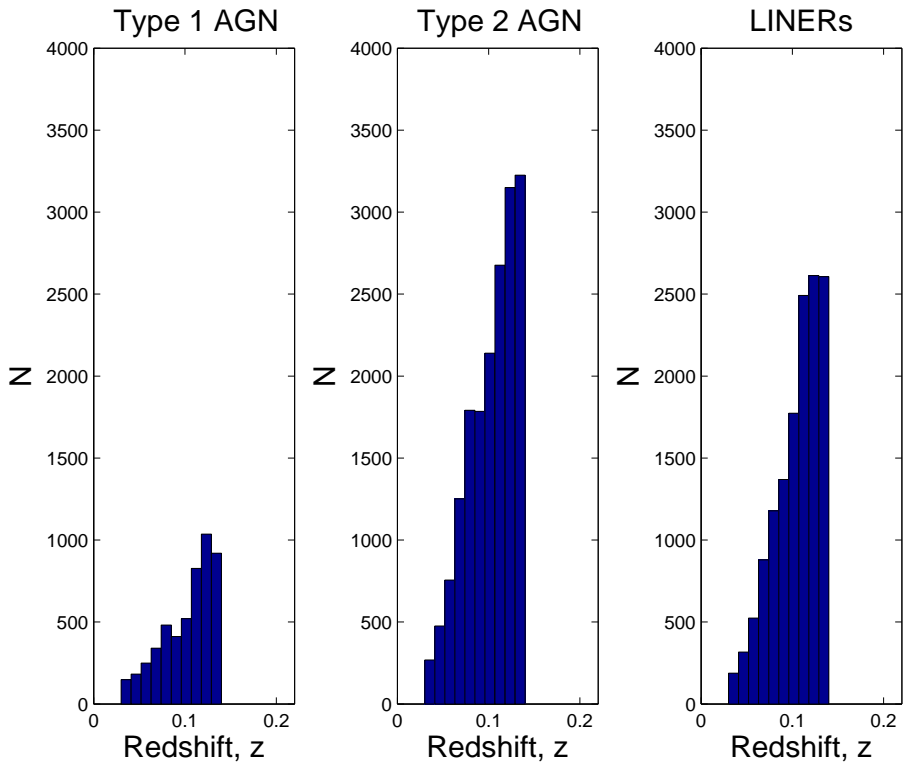}
     \caption{Distribution of redshift in the volume-limited parent AGN samples.}
               \label{ComparisionRedshift}
     \end{figure*}

We chose objects with M$_{r}$ $<$ - 21.2 and z $<$ 0.14 in order to get the largest possible number statistics (with respect to the number of Type 1 AGN) for a volume-limited subselection of our Type-1 AGN, Type-2 AGN and LINER Parent Samples. The redshift distributions of the three volume-limited subsample can be seen in Figure \ref{ComparisionRedshift}. It is in fair agreement with the redshift distribution of samples presented in Liu et al. (2011) of the QSO sample \citep{Schneider2010} and the narrow-line AGN sample selected with the Kauffmann et al.(2003) criterion.

\begin{figure*}
 \centering
  \includegraphics[scale=.7]{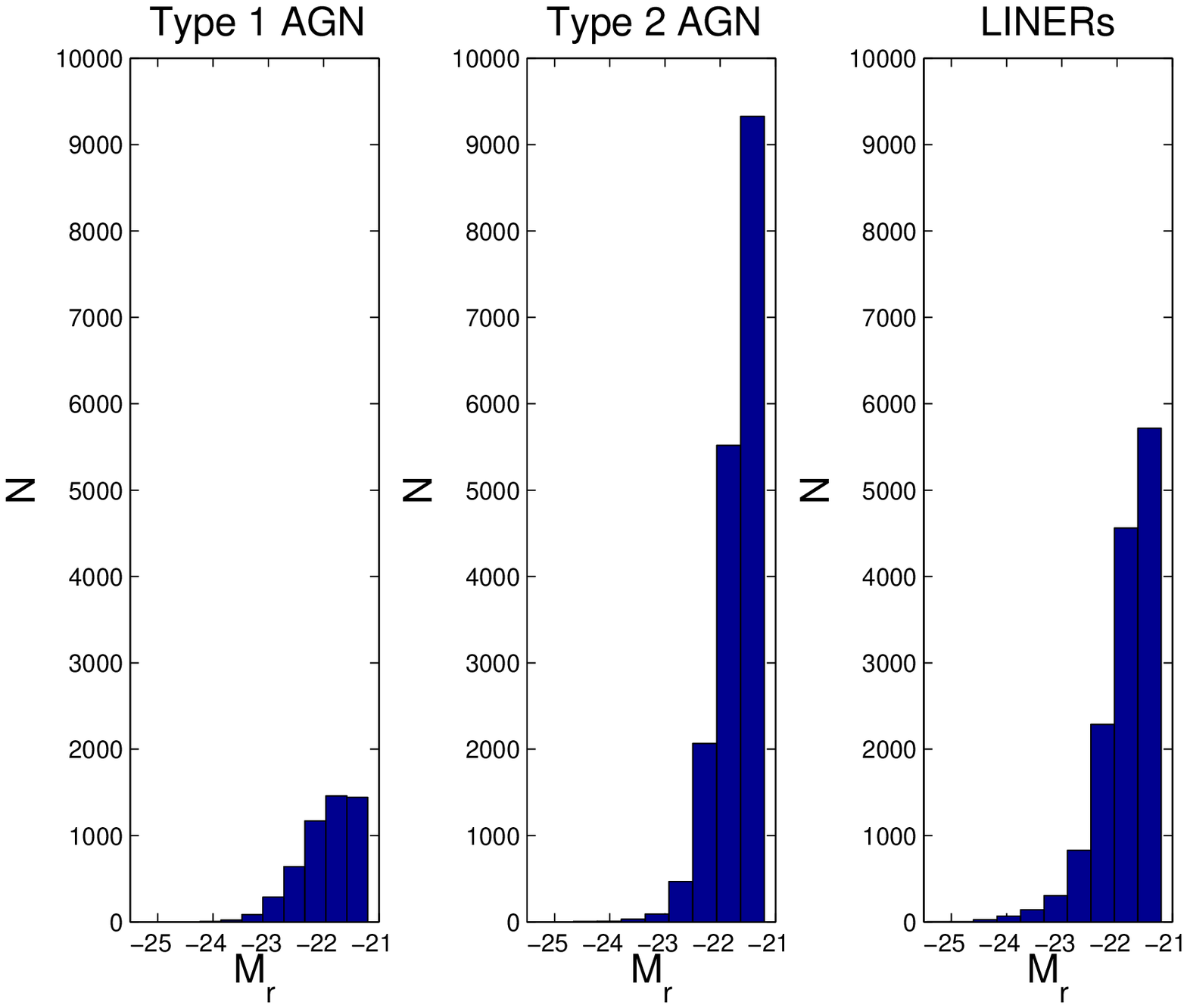}
     \caption{Distribution of $M_{r}$ in the volume-limited parent AGN samples.}
               \label{ComparisionMagnitude}
     \end{figure*}

Our parent samples suggests that we see approximately ~ 3.5 times as many Type-2 AGN as we have Type-1 AGN in this volume-limited subsample, see Table \ref{SumTables}. This is not in contradiction with other studies where the ratio for Type-1 to Type-2 AGN varies between 1:2 to 1:5 \citep[e.g.][]{Maia2003}. We also include
the absolute magnitude distribution for the same samples, figure \ref{ComparisionMagnitude}, from which
we cannot draw any conclusions in the light of Realistic Unification.

\subsubsection{Neighbours to AGN}

To select neighbours to our AGN in the parent samples, we query for all nearby galaxies from the SpecObjAll catalogue (DR7) to our AGN within 11 arcminutes with the function "dbo.fDistanceArcMinEq" and within $|\Delta$z$|<0.012 $ from the main AGN. This angular separation was selected in order to avoid a biased sample within the projected distance range of interest, which in our study is between 0 to 350 kpc. We are interested in this range as it covers both close interacting pairs as well as the nearby large-scale environment, where interesting things-- such as a sudden gap in the surface density of blue neighbours to quasars-- might be happening at a distance around 150 kpc \citep{Villarroel2012}. The redshift difference cut $|\Delta$z$|<0.012 $ is chosen at it provides us with the interesting opportunity to probe the large-scale environment in the redshift dimension, as we can investigate the clustering of galaxies by comparing the number of pairs within $|\Delta$z$| < 0.001$ (at z=0.2 corresponding to 4 Mpc), $|\Delta$z$| < 0.006$ (26 Mpc) and $|\Delta$z$| < 0.012$ (53 Mpc). Also, it has the benefit of being comparable to the photometric redshift neighbour sample as this value is roughly half of the photometric redshift error, $|\Delta$z$|<0.025$. For the photometric neighbour samples we later selected a redshift difference cut $|\Delta$z$|<0.03 $ (see following section). Together with the information from $|\Delta$z$| < $0.001, 0.006, 0.012 and 0.03 we cover the transition region between the spectroscopic and photometric neighbour samples and can thus discover potential effects in the neighbours caused by background galaxy contamination.

Due to spectroscopic fiber collisions approximately 2/3 of all pairs with angular separation less than 55$\arcsec$ can not be detected \citep[e.g.][]{Ellison2008}, unless residing on overlapping spectroscopic plug plates. This might bias our sample towards more wide pairs. We will, however, compensate the bias in the next section, when we will search for all remaining neighbours with photometric redshifts to our AGN in the spectroscopic redshift pair samples. The finite fiber sizes also makes it difficult to detect pairs in the late stages of mergers.

After removing repetitions, we retrieved 1658 Type-1 AGN-galaxy pairs (defined as the "Type 1 Spectroscopic Pair Sample"), 5698 Type-2 AGN-galaxy pairs ("Type 2 Spectroscopic Pair Sample") and 4214 LINER-galaxy pairs ("LINER Spectroscopic Pair Sample"). For these we obtain the dereddened apparent magnitudes, the spectroscopic redshifts, the isophotal axis lengths, the spectral line information, K-corrections from the PhotoZ catalogue and calculate the projected distances between the two object in each pair. We also calculate the rest-frame absolute magnitudes for each object as described in section 2.1.

\subsubsection{Photometric redshift neighbours}

To get a more valid estimate on the clustering of different colour types of galaxies around Type-1 and Type-2, we query from the PhotoZ catalogue to find all neighbour galaxies with photometric redshifts. We do this only for the AGN already existing in our spectroscopic AGN-galaxy pair catalogues and select all neighbours within the angular distance of 11 arcmin and and $|\Delta$z$|<0.03 $ from the AGN.

While the photometric redshift samples do not suffer from the fiber incompleteness issue, the resolution and surface brightness limits of the SDSS photometry might play a role in the detection of neighbours. Faint neighbours might not be detected, and if any AGN has an increased number of low surface brightness (LSB) galaxies at short projected separations, these might not be observed and hence
bias our sample towards an underestimated clustering of LSB galaxies around the AGN. For the photometric pairs, we retreive the morphologies for all AGN hosts from the Galaxy Zoo project, see below. Our final samples (including morphologies) are 13519 Type-1 AGN-galaxy pairs (defined as "Type 1 Photometric Pair Sample") and 58743 Type-2 AGN-galaxy pairs (Type 2 "Photometric Pair Sample"). The ratios of pairs between the photometric pair samples and spectoscopic pair samples indicate that the Type-2 AGN have more nearby neighbours than what can be detected by SDSS spectroscopy.

\subsubsection{Morphology samples}

For the pairs, we chose to query for the morphology of the parent, AGN hosts in the pair from the Galaxy Zoo\footnote{The SDSS casjobs interface allows one to obtain the
Galaxy Zoo classification for any object in their catalog (Table 2) with a corresponding SDSS Object ID.} project \citep{Lintott,Lintott2010}. Not all parent AGN in our pair samples had morphology information. This is why the morphology samples are very similar to, but slightly smaller, than the spectroscopic redshift samples. They are defined as "Morphology Pair Samples". We also obtain the morphology classifications for our parent AGN samples, so that we can see if the presence of a close neighbour significantly changes the morphology of the AGN hosts, defined as "Morphology Parent Samples".

\begin{table*}[ht]
\caption{A summary of all the samples in our study. The number of objects in the volume-limited cut of the sample is given (z $<$ 0.14 and M$_{r}<-21.2$). The total number of objects in each sample is given in parenthesis.}
\centering
\begin{tabular}{c c c c}

\multicolumn{4}{c}{Samples} \\
\hline\hline
Sample type & Type 1 & Type 2 & LINER\\ [0.2ex]
\hline
Parent & 5114 (11334) & 17518 (53416) & 13939 (34249) \\ 	
Spectroscopic Pair & 703 (1658) & 1194 (5698) & 1282 (4214) \\ 	 
Morphology Parent & 5065 (10154) & 16294 (48970) & - \\ 
Morphology Pair & 702 (1604) & 1191 (5547) & -\\ 
Photometric Pair & 1836 (13519) & 3663 (58743) & - \\[0.2ex]
\hline
\end{tabular}
\label{SumTables}
\end{table*}

\subsection{Line flux and extinction corrections}

For the neighbouring galaxies to the AGN in our spectroscopic redshift pairs, we correct the Balmer emission line fluxes and equivalent widths for underlying stellar absorption by assuming average absorption line strengths corresponding to 2.5 \AA\ in $EW$ for H$\alpha$ and 4 \AA\ for H$\beta$ \citep{Mattsson2005}. Further, we do internal extinction corrections for galaxies with $EW$(H$\beta$) $>$ 5 \AA\ according to a standard interstellar extinction curve \citep[e.g.][]{Osterbrock2009,Whitford}. Here the stellar absorption-corrected H$\alpha$ and H$\beta$ lines for each galaxy are used to calculate the extinction coefficient, that together with the interstellar extinction curve can be used to estimate the amount of extinction for each emission line.

The colours of the neighbours were corrected for inclination-dependent dust extinction, using the analytical expressions in Cho \& Park (2009), meaning that we corrected the colours for all galaxies having 0 $<$ (u-r) $<$ 4, H$\alpha$ line width ranging from 0 to 200 \AA\, and absolute magnitude - 21.95 $ < M_{r} < $ - 19.95 and finally a concentration index C in the range 1.74 $< C <$ 3.06. Neighbour galaxies outside these parameter ranges are left untreated, which for instance means that more luminous objects have no inclination-dependent extinction correction on their colours applied. Unless all neighbour galaxies around only one certain type of AGN would be observed with the same inclination angle (which is highly unlikely due to the assumed isotropy of the Universe) $or$ the selection of the two AGN types has strong biases, we do not expect extinction correction to significantly change the results. We control this by not including any extinction corrections on colours and emission lines, which turns out not to change the results in the paper at all and the applied extinction corrections are therefore merely used for improved accuracy.

\section{Results}

\subsection{Volume-limited test sample of pairs}

We estimate the relative fraction of pairs by comparing the spectroscopic pair samples to the spectroscopic parent samples.

\begin{table*}[ht]
\caption{Number of satellites around central AGN in a volume-limited pair samples. The number
of satellites in the subsample $|\Delta z|$=0.001 (column 2) is compared to the Parent Sample $|\Delta z|$=0.012 (column 4). 
The fraction of satellites in the subsample $|\Delta z|$=0.001 (column 3). The ratio (column 5) between the number of satellites in subsample $|\Delta z|$=0.001 and total number of objects in a volume-limited
pair sample (column 6).}
\centering
\begin{tabular}{c c c c c c}
\multicolumn{6}{c}{Clustering of neighbours} \\
\hline\hline
Central galaxy type & $|\Delta z|$=0.001 & fraction$_{\Delta z}$ & $|\Delta z|$=0.012 & fraction$_{pairs}$ & N Parent Sample \\ [0.2ex]
\hline
Type-1 & 385 & 0.55 & 703 & 0.08 & 5114\\ 
Type-2 & 718 & 0.60 & 1194 & 0.04 & 17518 \\ 
LINER & 794 & 0.62 & 1282 & 0.06 & 13939\\ [0.2ex]
\hline
\end{tabular}
\label{Sample}
\end{table*}

From the second column of Table \ref{Sample}, we can not see any significant difference 
in $|\Delta z|$ between the different pair samples. On the other hand, we do see that the Type-1 AGN significantly more often have a close neighbour than do Type-2 AGN or LINERs. For the AGN themselves we also compare the absolute magnitude $M_{r}$, rest-frame colour $u_{e}-r_{e}$ and redshift distributions to AGN in the parent samples. No significant changes in luminosity, colour and redshift, depending on the presence or absence of a companion, are found.

\subsection{Colours of neighbours}

Dust extinction, metallicity and the age distribution of stellar populations are all factors that can influence the colour of a galaxy. We therefore investigate how the colour $u_{e}-r_{e}$ of a neighbour galaxy to AGN changes as a function of distance from the AGN.

\begin{figure}
 \centering
  \includegraphics[width=8cm]{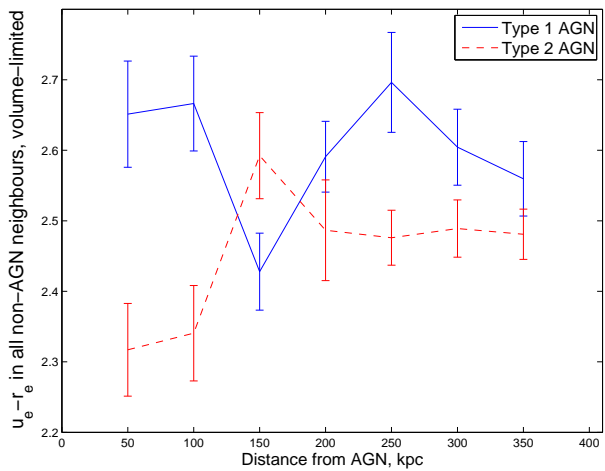}
     \caption{Volume-limited comparison of internal extinction corrected $u_{e}-r_{e}$ colour of non-AGN neighbour galaxies versus distance to the AGN with $|\Delta z|$ $<$ 0.012. The number of pairs in the plot is 582 for the Type-1 AGN and 913 for the Type-2 AGN. Among the neighbours, all AGN, LINERs and galaxies with exceptional colours $u_{e}-r_{e}$ $>$ 10 are removed.}
               \label{VolumeColor}
     \end{figure}

Figure \ref{VolumeColor} shows how the internal extinction corrected $u_{e}-r_{e}$ colour \footnote{The suffix "e" means "internal 
extinction corrected" in this context.} for a normal (non-AGN) companion galaxy changes as a function of the projected distance between the companion and the AGN in bins of 50 kpc. Standard errors are indicated by the 1$\sigma$ error bars.

\begin{figure}
 \centering
  \includegraphics[width=8cm]{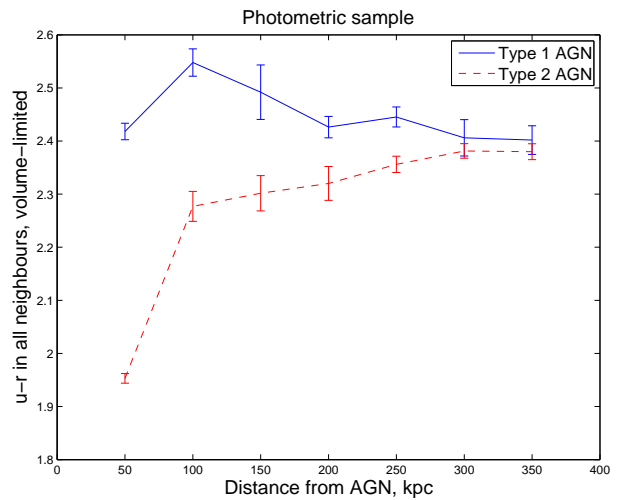}
     \caption{Volume-limited comparison of $u-r$ colour of neighbour galaxies versus distance to the AGN with $|\Delta z|$ $<$ 0.03
from the photometric samples. The number of pairs in the plot is 1836 for the Type-1 AGN and 3662 for the Type-2 AGN.}
               \label{VolumeColorPhoto}
     \end{figure}

The average colour of the neighbours is redder around Type-1 AGN than around Type-2 AGN.
We redo the analysis by also using two more $|\Delta z|$ cuts: $|\Delta z|$ $<$ 0.001 and $|\Delta z|$ $<$ 0.006 and observe the same differences. Contamination of background galaxies is therefore likely not to be the cause. Since the loss of an increased number of low-luminosity neighbours at small projected separations from fiber collision is larger for pairs at higher redshift (as the Type-1 AGN-neighbour pairs), we do a similar analysis with our photometric neighbour samples, see Figure \ref{VolumeColorPhoto}. The colours of the neighbours from the photometric pair samples are not corrected for internal extinction. We can see that the colours of the neighbours
in the photometric pair samples are bluer than those in the spectroscopic pair samples. This is a consequence
of that we have not removed the AGN among the photometric redshift neighbours as we lack emission line information for these, which means that we include a significant fraction of blue, star-forming galaxies in Figure \ref{VolumeColorPhoto}.

The photometric samples show that the combined effects of fiber collisions and spectroscopic targeting limit cause an underestimation of the number of neighbours, something that is clearer reflected in the Type-2 AGN neighbour samples. While the ratio of the pairs in the volume-limited cuts equals to 
spectroscopic/photometric $\sim$ 0.38 for the Type-1 morphology neighbour samples, the same ratio for Type-2 is $\sim$ 0.33. We do the same plots for three types of morphological hosts. The volume-limited cuts of the Spiral and Uncertain samples of Type-1 and Type-2 AGN hosts exhibits exactly the same trend as does Fig \ref{VolumeColorPhoto}. For the Elliptical hosts, we had too few Type-2 neighbours to do a similar analysis, but for the closest bin within 50 kpc we calculated the mean value for Type-2 neighbours ($u-r$ $\sim$ 2.02 $\pm$ 0.04) and for the Type-1 AGN neighbours ($u-r$ $\sim$ 2.53 $\pm$ 0.01). Possible restrictions of AGN unification to a certain galaxy morphology is thus not sufficient to explain the differences between the two AGN populations.

\subsection{Star formation rate}

We use the measured H$\alpha$\ emission line from the spectroscopic catalog
of the SDSS together with the Bergvall-R\"onnback calibration (1995) to estimate the star-formation rate in the neighbours of AGN. 

\begin{equation}\label{eq4}
L(H\alpha)= SFR *1.51*10^{34}
\end{equation}
and

\begin{equation}\label{eq5}
L(H\alpha)=4 \pi D_L^{2} \sqrt{2\pi} \sigma h 10^{-20}
\end{equation}

where SFR is the star formation rate in solar masses per year, $\sigma$ and $h$ are width and 
height of the H$\alpha$ emission line, $D_L$ is the luminosity distance in Mpc and the emission line luminosity is expressed in watts.

We first try to see if we can find any correlation or differences of SFR with distance between AGN and neighbour within the volume-limited cuts. The small number of Type-1-galaxy pairs (153) and Type-2-galaxy pairs (365) within the volume-limited cut does not permit us to detect any statistically significant differences in the neighbour samples. We therefore search for an indication by estimating the number of neighbours (see Table \ref{SampleSFRAGN}) within this volume-limited cut that have measurable star-formation \footnote{In our study, we define the star-formation rate as "measurable" if the flux in H$\alpha$ $>$ 0 and the flux is twice the flux error.} We find that 31 \% of the Type-2 AGN neighbours have measurable star-formation rate, while only 22 \% of the Type-1 AGN neighbours, with the statistical significance level of 3.3$\sigma$. This could suggests that the Type-2 AGN either form in regions with abundant gas supply or that they could transfer gas to the neighbour galaxies. Any of these explanations would support the observed higher column densities of molecular hydrogen in the Type-2 AGN \citep{Awaki1991}. One could also argue that Type-2 AGN are dustier objects on average, and therefore could suppress the star-formation rate of the neighbours less. We in addition notice that the Type-1 AGN neighbours have a higher dust content (measured in H$\alpha$/H$\beta$-ratio) than the Type-2 neighbours, but since the H$\alpha$ line is corrected for underlying stellar absorption, Galactic and internal extinction, the dust extinction is suggested not to be the cause of the observed differences of star-formation rates.

\begin{table*}[ht]
\centering
\caption{The number of neighbours with measurable SFR, defined as blue and defined as AGN. $|\Delta z|$=0.012 is used.}
\begin{tabular}{c c c c c c c c}
\multicolumn{8}{c}{Blue \& AGN neighbours} \\
\hline\hline
Central galaxy type & measurable SFR & fraction$_{SFR}$ & blue & fraction$_{blue}$ & AGN & fraction$_{AGN}$ & Total number of pairs \\ [0.2ex]
\hline
Type-1 & 412 & 0.25 & 527 & 0.32 & 243 & 0.15 & 1658\\
- volume-limited & 153 & 0.22 & 119 & 0.17 & 110 & 0.16 & 703\\ 
Type-2 & 2228 & 0.39 & 2323 & 0.41 & 1001 & 0.18 & 5698\\ 
- volume-limited & 365 & 0.31 & 265 & 0.22 & 238 & 0.20 & 1194\\ [0.2ex]
\hline
\end{tabular}
\label{SampleSFRAGN}
\end{table*}


\subsection{Overdensities around Type-1 and Type-2 AGN}

\begin{figure}
 \centering
\caption{Ratio of Type1/Type2 AGN companions around Type-2 AGN as a function
of distance from the Type-2 AGN. LINERs are excluded and we use neighbours with $|\Delta z|$ $<$ 0.012. 
In total, there are 92 Type-1 AGN companions and 527 Type-2 AGN companions in the plot.}
  \includegraphics[width=8cm]{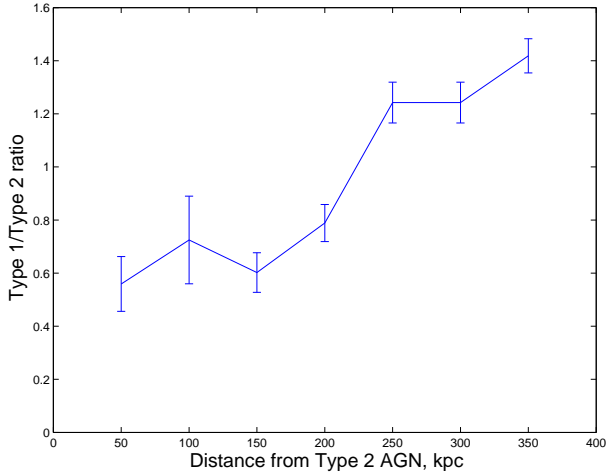}
               \label{KvotArp}
     \end{figure}

We are interested to see how the surface densities of AGN neighbours to Type-1 and Type-2 are interconnected to understand the differences in the small-scale environment around different types of AGN. We calculate the annular surface densities by

\begin{equation}
\rho=N_{\rm distance}/((d_{1})^2-(d_{2})^2)\pi
\end{equation}\label{eq3}

where $N$ is the number of companion galaxies in each annulus with outer radius $d_{1}$ and inner radius $d_{2}$. The companion galaxies are binned into ten pieces of 35 kpc each, and we calculate the Poisson errors
for each bin. Further, we calculate how the ratio of Type-1/Type-2 neighbours around Type-2 AGN varies as a function of distance from the Type-2 AGN with accompanying error estimation, presented by the 1$\sigma$ error bars, see Figure \ref{KvotArp}. We get a clear 
decrease (with 4.5 $\sigma$ significance) of the ratio at close projected separations. We see the same trend also when using $|\Delta z|$ $<$ 0.001. As this is a ratio, possible biases in the ratio will tend to cancel out each other. We thus argue that this is an actual physical difference between the two main classes of AGN. It could explain the lack of observed Type-1 AGN in isolated pairs of galaxies by Gonzalez et al. (2008).

\subsection{Morphology}

The morphology of AGN hosts is a long-disputed question and we here approach it by comparing the morphologies of Type-1 AGN and Type-2 AGN in the parent samples (Table \ref{MorphologyHostControlSample}) with those in the pair samples (Table \ref{MorphologyNeighbours}). In this way, we will be able to see how the detection of a nearby neighbour could influence the observed morphology of the AGN host. An interesting feature is that while Type-1 AGN having a close neighbour tend to be more elliptical in the parent sample (something that indicates a merger transforming the spiral host), Type-2 AGN always reside in spiral hosts. The more elliptical form of Type-1 AGN hosts in pair could also be a result of being in denser clusters. Our clustering (Table \ref{Sample}) shows that the Type-1 AGN actually reside in less dense clusters than do Type-2 AGN. The lack of Type-2 AGN in elliptical hosts suggests that the state of a Type-2 AGN might be extremely shortlived in comparison to Type-1 AGN, and that a merger might convert the Type-2 AGN into a different object, possibly a Type-1 AGN or a LINER. The reason why we suggest this is that the presence of a neighbour can be seen as an indicator of a denser environment where the probability of merger is larger. It is commonly believed that a merger will transform a spiral galaxy into an elliptical, which is also the case we observe for the Type-1 AGN where the presence of a neighbour significantly decreases the fraction of host galaxies with a spiral morphology and increases the fraction of galaxies with elliptical morphology. This is however not the case for Type-2 AGN, for which we cannot see any increase at all in the fraction of elliptical host in the presence of a close companion. The dramatic difference in the two cases would imply two of the following options: either spiral galaxy hosting a Type-2 AGN does not get influenced by a merger or close companion (by some exotic mechanism), or the Type-2 AGN cannot be preserved in its “Type-2” state after a merger. The latter explanation seems to be more in picture with the observed the lack of Type-2 AGN (and presence of Type-1 AGN) in the high-redshift Universe, where the merger rate is significantly larger. We additionally compare the fraction of LINERs in elliptical galaxies in the parent sample (2.5 \%) and in a similar sample of LINERs with neighbours (5.3 \%), and conclude that either the Type-2 AGN only exist in spiral form before the merger or the Type-2 AGN cannot merge with their neighbours.

\begin{table*}[ht]
\caption{Parent sample. Morphologies of Type-1 and Type-2 AGN in volume-limited parent samples consisting of all Type-1 and Type-2 AGN within this volume, disregarding if they have neighbour or not. Errors in redshift are less than $\delta_{z}$ $<$ 0.01.}
\centering
\begin{tabular}{c c c c c c}
\multicolumn{6}{c}{Morphology} \\
\hline\hline
Host & spiral & z spiral host & elliptical & z elliptical host & total\\ [0.2ex]
\hline
Type-1 & 1128 (22\%) & 0.123 & 751 (15\%) & 0.123 & 5065\\
Type-2 & 8506 (52\%) & 0.098 & 81 ($<$1\%) & 0.098 & 16294\\[0.2ex]
\hline
\end{tabular}
\label{MorphologyHostControlSample}
\end{table*}

\begin{table*}[ht]
\centering
\caption{The morphology of host galaxy with neighbours in a volume-limited sample. Not all galaxies had morphologies from the Galaxy Zoo, and those with uncertain ones are not included in the table. Errors in redshift are $\delta_{z}$ $\sim$ 0.01.}

\begin{tabular}{c c c c c c}
\hline\hline
Central galaxy type & spiral hosts & z spiral host & elliptical hosts & z elliptical host & total number of hosts\\ [0.2ex]
\hline
Type-1 & 113 (16\%) & 0.098 & 145 (21\%) & 0.087 & 702\\
Type-2 & 676 (57\%) & 0.095 & 25 (2\%) & 0.072 & 1191\\[0.2ex]
\hline
\end{tabular}
\label{MorphologyNeighbours}
\end{table*}

\subsection{A Hypothetical Luminosity Test of AGN unification}

As the intrinsic properties of AGN are unknown due to the lack of knowledge on the geometry of the dust torus, the high discrepancies between the neighbour populations of Type-1 and Type-2 AGN could be the result of a biased selection on bolometric luminosities and masses of the objects. It is very difficult to estimate how much the torus on average obscures the Type-2 AGN in the case of Realistic AGN unification due to the little observational evidence. The only thing we know is that if an AGN would be obscured by a dust torus, it would appear less luminous and could be mistaken for being less massive. This means that Type-2 AGN of lower luminosities could be considerably more massive than Type-1 AGN of the same luminosities.

One obvious way to approach the problem of a possible mass bias would be to extract the stellar and black hole masses of the central AGN in the two samples. We, however, will do this by matching the neighbour samples in colour, while estimating the luminosity displacement in the Type-2 AGN relative to the Type-1 AGN, assuming a Gaussian distribution of the covering factor.

The risk of doing this for random two samples of Type-1 and Type-2 AGN is that in case the accretion disk of the Type-1 AGN outshines the host galaxy, one could end up comparing the luminosity from the accretion disk in Type-1 AGN to the luminosity of the host galaxy in Type-2 AGN. The advantage of using low-redshift AGN for this study becomes clear in the context; the assumingly equal contribution (under the assumption of AGN unification) of the host galaxy luminosity in rather faint low-redshift AGN makes it possible to compare the luminosity displacement on a statistical level, without any concerns for that the accretion disk of the Type-1 AGN could outshine the the host galaxy as in high-redshift AGN. As a result we are able as a check of consistency, to investigate if there exists any luminosity displacement where the neighbours of the two classes of AGN have the similar neighbours (and thus assumingly the similar masses), and if they further have the same distributions of morphologies among host galaxies, which is a necessary condition for AGN unification to be true.

We herein will propose a test to sort out whether a luminosity displacement is enough to explain the discrepancies.

\subsubsection{Step one: determining the variables}

While the general clustering of galaxies around the central AGN is rather similar for the two populations, we have seen that the pairs mainly differ in the following properties:

\begin{enumerate}

\item Clustering of galaxies with measurable star-formation around the central AGN.
\item Average colour of the neighbour population.
\item Correlation of colour with distance between neighbour and AGN.
\item Morphology of the AGN host galaxy.

\end{enumerate}

We here define the luminosity displacement in the AGN, E$_{dis}$

\begin{equation}
\Delta E_{dis} = M_{r,tot}-M_{r}
\end{equation}

We define an average "offset" (O$_{colour}$) in the colour $u-r$ between the neighbour populations of the Type-1 and Type-2 populations for projected distances d $>$ 50 kpc. The distance where the neighbour is no longer under direct influence of the AGN is estimated from Figure \ref{VolumeColorPhoto}. The reason for this choice is that a potential Holmberg effect \citep{Holmberg1969}, the possibility that more star-forming galaxy neighbours tend to be aligned perpendicular to the disk of the galaxy, could influence the star formation rate at short separations (d $<$ 50 kpc), but would be unnoticed at larger radii. However, in this particular case, we would also see the same distribution of AGN host morphologies among Type-1 and the luminosity-displaced Type-2 AGN.

\subsubsection{Step two: starting sample}

The starting point consists of two volume-limited subsamples of both Type-1 and Type-2 AGN, where $M_{r}$ $<$ - 21.2 and 
z $<$ 0.14 is used for both central AGN and neighbour galaxies. While $M_{r}$ will be held constant for the neighbour galaxies and the Type-1 AGN, we will vary this parameter for the central Type-2 AGN until we reach a state where the offset will be close to O$_{colour}$ $\sim$ 0 and both neighbour populations will be very similar to each other. We define this point as the first $M_{r}$ value where the 1 $\sigma$ error bars of the two sample overlap.

We will perform the test for the two extreme cases to probe possible hidden obscured Type-1 AGN among the Type-2 AGN, the first as narrow as the volume limited Type-1 AGN sample in magnitude ("Single magnitude cut"), the second infinitely broad ("Double magnitude cuts"):

\begin{enumerate}
\item Varying the lower luminosity cut $M_{r}$ $<$ - 21.2, where we iterate it in the direction $M_{r}$ $>$ - 21.2 as far as we have data.
\item Varying the lower luminosity cut $M_{r}$ $<$ - 21.2, where we iterate it in the direction $M_{r}$ $>$ - 21.2 as far as we have data, plus an upper cut $M_{r,upper}$ where the range is equivalent to the fix width of the volume-limited distribution of Type-1 AGN in the pair sample. Since the most luminous Type-1 AGN in our volume-limited pair sample have an absolute magnitude $M_{rmax,Type-1 AGN}$= - 23.8, this gives us a comoving, fixed width of w=2.6 mag.
\end{enumerate}

A support for AGN unification would occur during the iterations if O$_{colour}$ $\sim$ 0, and the new, deduced correlation of and colour (Figure \ref{VolumeColorPhoto}) with projected distance are fairly similar between the two populations and the Type-2 AGN pair subsample with O$_{colour}$ $\sim$ 0. A second condition is that the new suggested Type-2 luminosity population must have roughly equal fractions of spiral and elliptical AGN among the hosts.

The average redshift of two populations with similar offsets also can reveal a minor time-evolution that otherwise can go unnoticed. The same test will also be used in future publications to connect LINERs to the evolution of active galaxies.

\subsubsection{Test results}

\begin{figure}
 \centering
\caption{Results of the hypothetical luminosity test. The x-axis shows the lower cut in absolute magnitude 
both for the one-cut and the fixed-width luminosity test. Y-axis shows the K- and Galactic extinction corrected colours among the neighbours.}
  \includegraphics[width=8cm]{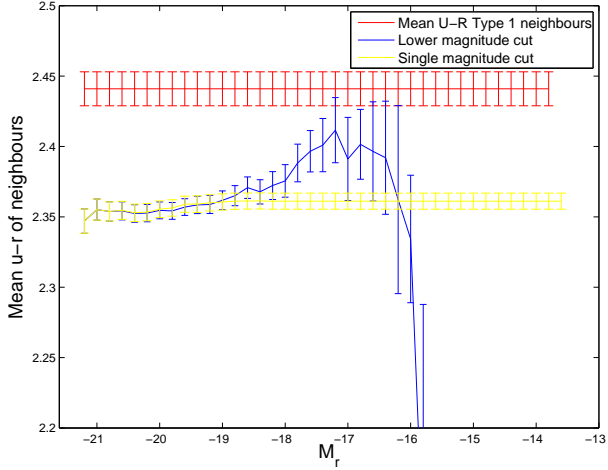}
               \label{Test}
     \end{figure}

Figure \ref{Test} shows the results indicating that a subsample of Type-2 AGN within the luminosity range of - 17.2 $<$ $M_{r,Type-2 AGN}$ $<$ -19.8 (corresponding to a luminosity displacement E$_{dis}$ $\sim$ 4 mag) could correspond in mass to the volume-limited Type-1 in the pair samples. Using only a lower cut did not influence the selection. The next step was to redo the colour-distance plot (Figure \ref{VolumeColorPhoto}) with these luminosity constraints imposed on the Type-2 host luminosity selection and investigate the relative fractions of the host morphologies within. The plot revealed that if selecting Type-2 AGN with the magnitude range - 17.2 $<$ $M_{r,Type-2 AGN}$ $<$ -19.8, the Type-2 AGN neighbours turn out to be very similar to the Type-1 AGN neighbours in their colour-distance behavior. However, the distribution of morphologies were similar to those in the volume-limited parent sample, see Table \ref{MorphologyHostControlSample}. In this selection among the Type-2 AGN with neighbours, 25\% could be detected in spiral hosts and only 2\% in elliptical hosts, while for the parent sample the corresponding number was 52\% in spiral hosts and 1.5\% in elliptical. Even if the different covering factors of the torus are insufficient to explain the observed difference in morphology populations, the similarity in neighbours of Type-1 AGN with more faint Type-2 AGN suggests a strong evolutionary link between the two types of AGN, where the change of morphology during interaction seems to be one of the key ingredients.

We also redid this test with the sample of Spiral AGN, but here no magnitude range or cut could set
the colour offset O$_{colour}$) to zero.

Additionally, we investigated how the S/N ratio in any of the diagnostic lines to select the AGN could influence our results by following the same principle as above but with S/N ratio on the x-axis and the same magnitude
limits for both samples $M_{r}$ $<$ - 21.2 (both for AGN and its neighbour). We did this for all emission lines separately for both Type-2 AGN and the Type-1 AGN's H$\alpha$ emission line (including the width), but also for all diagnostic lines at the same time. We also include a similar requirement in S/N-width on both H$\alpha$ and the [\ion{N}{ii}] as the single-Gaussian selection mode of Type-1 could have been the root of the observed differences, and by chance could result in the inclusion of non-AGN with poorly resolved H$\alpha$ and [\ion{N}{ii}]6585. However, increasing S/N ratio did not set the O$_{colour}$ $\sim$ 0, and hence the emission line diagnostics is unlikely to cause any significant bias in our results.

\section{Discussion and conclusions}

We construct two parent samples of broad-line (Type-2) and narrow-line (Type-1) AGN at redshifts 
0.03 $<$ z $<$ 0.2 from the SDSS DR7 by using H$\alpha$ emission line width and BPT-diagrams combined with \cite{Kauffmann2003} criteria. For the objects in the parent samples, we select spectroscopic neighbours with the redshift difference cut $|\Delta$z$|<0.012 $ and within the projected distance of 350 kpc.
From the Photo-z catalogue, we also create a comparison sample with neighbours having photometric redshifts and a redshift difference $|\Delta$z$|<0.03 $. For all AGN hosts
and neighbours in the spectroscopic samples, we retrieve morphology classification from Galaxy Zoo, K-corrected absolute magnitudes and spectral line measurements. We identify the AGN among the neighbours to our Type-1 and Type-2 AGN. The spectral line measurement of the non-AGN neighbours are corrected for underlying stellar absorption and internal dust extinction.

For the neighbours, we measure the colours and star-formation rates as a function of distance in volume-limited cuts of the samples, where both neighbour and hosts have $M_{r}$ $<$ - 21.2.

The main conclusions are:

\begin{enumerate}

\item Type-2 AGN neighbours are on average bluer than Type-1 AGN neighbours.
\item The colour in a non-AGN neighbour to Type-2 AGN changes dramatically at short separations, while the colour in the Type-1 neighbour remains uninfluenced. 
\item The fraction of star-forming neighbours around Type-2 AGN is higher than around Type-1 AGN.
\item The ratio of Type1/Type2 AGN neighbours to Type-2 AGN increases with separation. The correlation has a statistical significance of 4.5 $\sigma$.
\item The morphology of Type-2 AGN is uninfluenced by a close companion (contrary to the case of Type-1 AGN). Either Type-2 AGN rarely merge, or the Type-2 AGN is an extremely short-lived AGN state.
\item The differences between the complete parent samples and the volume-limited parent samples suggest
that Type-2 AGN are surrounded by many more dwarf galaxies than are Type-1 AGN.
\item Using AGN with only a certain type of morphology (spiral or elliptical) when investigating their
influence on the neighbours does not change the above-mentioned results.
\item Approximating a value on the luminosity displacement between Type-1 and Type-2 AGN needed to reproduce a similar colour distribution of neighbours around the two types yields a luminosity displacement E$_{dis}$ $\sim$ 4 mag, but can not explain the discrepancies in the morphology populations among the two AGN types. Thus, the Realistic Unification is insufficient to explain the difference between the two neighbour masses.
\item Increasing the S/N ratio of the diagnostic lines used in classification does not influence the results.

\end{enumerate}

From our results, we conclude that viewing angle is insufficient to explain the differences between the two neighbour samples independently of the nature of the dust torus.

\subsection{Possible biases}

At this point, one should question whether the different environments of Type-1 and Type-2 AGN might be caused by various kinds of selection effects.

The most important biases might come from the sample selection itself. While we used quite convential methods for creating the samples with the help of the Kauffmann (2003) criterion for the Type-2 AGN with and $\sigma$(H$\alpha$) $<$ 10 \AA\ , the criteria used for the Type-1 AGN was single-Gaussian emission-line width $\sigma$(H$\alpha$) $>$ 10 \AA\ and thus depends on a single emission line measurement. The difficulties of resolving H$\alpha$6565 and [\ion{N}{ii}]6585 lines could cause a bias and one should possibly rather use multi-Gaussian fits for selecting broad-line AGN. Our approach to the problem was to examine if increasing the S/N ratio of the emission line fluxes and widths for both for the H$\alpha$6565 and [\ion{N}{ii}]6585 emission lines would change the differences between the two samples in any way. If the choice of single-Gaussian line fit is the cause of selecting fake broad-line AGN, increased demands on the S/N in both emission line widths and fluxes should decrease the number of potential fake broad-line AGN. However, no changes were observed by introducing this criterion. Also, at earlier stages of the analysis we used $\sigma$(H$\alpha$)= 15 \AA\ as a limit, which yielded the same results. In a future publication, we will do the same analysis using more sophisticated broad-line samples \citep[e.g.][]{Hao2005,Schneider2010} and narrow-line 
samples \citep[e.g.][]{Reyes2008,Aihara2011} as well as using the [\ion{O}{ii}] emission line for classification. The advantage of using these is that they also take into account contributions from stellar absorption and internal extinction from the host galaxy when constructing the samples, the neglect of which can be another bias in our sample selection. However, if AGN unification is true, there is no reason to expect that we would have different dust content or star-formation in the host galaxy itself, which is why such a bias should not alter our conclusions. Our sample ratios of Type-1 to Type-2 AGN in our volume-limited cuts is similar to that found by several the other groups \citep[e.g.][]{Hao2005,Maia2003}, 1:2, which supports that our sample selection methods are robust.

Also the sky-covering factor could be a cause of seeing different neighbours around Type-1 and Type-2 AGN. However, the large differences of morphologies of the AGN hosts rules out this possibility. Also, redshift and luminosity biases can be ruled out as they would not produce a correlation in the ratio Type1/Type2 neighbours around Type-2 as in Figure \ref{KvotArp}. The lack of observations of Type-1 AGN in isolated pairs of galaxies \citep{Gonzalez2008} supports this finding.

Another bias we considered is whether the light from a Type-2 AGN would contaminate the closest neighbours (the bin at the shortest projected separation) and therefore create a false trend in the colours. However, we examine the luminosity distributions of AGN hosts and note that the Type-2 AGN are not any more luminous than the Type-1 AGN thus light contamination can not be the cause of the differences in the closest bin. To check if background galaxies might be the cause, we also do all the plots using spectroscopic pairs in three different redshift cuts $|\Delta$z$|<0.001$, $|\Delta$z$|<0.006$ and $|\Delta$z$|<0.012$, which all yield the same results.

In our study we have not separated radio-quiet from radio-loud objects. Such a study would require much higher-precision radio measurements on all the AGN hosts than available in the catalogues today. A bias from including more radio-loud objects into one of the samples could also influence our results and conclusions. In future we hope to do this with a better categorization of a large number of objects. However, in the case that AGN unification were true, there is no reason to expect that we would have a different fraction of radio-loud objects in the two samples.

Finally, if the Realistic Unification were real, a clumpy dust torus might cause us to select AGN from different parts of the luminosity distribution and thus different stellar masses, and therefore different clustering of neighbours. We approached the problem by our hypothetical luminosity test and even if performing the hypothetical luminosity test on host galaxies initially (by only investigating the colours of neighbours) suggests a luminosity displacement corresponding to E$_{dis}$ $\sim$ 4 mag in Type-2 AGN, this is not the cause. The morphology differences of AGN host galaxies between the two population are in contradiction with the idea of Realistic Unification irrespective of whether we select Type-1 and Type-2 AGN with "colour-matched" neighbour populations. The strong trend in colour of Type-2 AGN might therefore be related to the observed higher column densities of hydrogen gas in the Type-2 AGN \citep{Awaki1991}, and would suggest a higher transfer of fuel into the Type-2 AGN neighbour. However, redoing this test using stellar masses of the host galaxies is needed to provide further support for our statement. Improved statistics on neighbour galaxies with measurable star-formation rate would also be needed to confirm (or reject) this hypothesis. 

\subsection{Time-evolved AGN unification?}

As we could not find any support for the notion that Type-1 and Type-2 AGN are intrinsically the same type of objects, we will consider the alternative of a time evolution between the two types of objects as suggested by the Krongold-Koulouridis scenario \citep{Koulouridis2006a}, where the interaction between gas-rich galaxies ignites starburst activity. The ignition of star-formation activity by interaction is something that has been confirmed by Ellison et al. (2010) in studies of galaxy pair colours and star-formation rate. In the time-evolved model, the starburst in the galaxy channels gas down to feed the super-massive black hole, and the accretion disk starts shining. As the starburst dies off, the accretion disk luminosity becomes more dominant and the remaining molecular gas and dust forms a torus around the disk. As the interaction proceeds, the AGN gets stronger and finally drives away the obscuring clouds, forming a Type-1 AGN.

The Krongold-Koulouridis scenario has several features that are consistent with our observations: 
the decrease of the Type1/Type2 AGN ratio near Type-2 AGN (assuming the Type-2 AGN formed in both galaxies
in the pair simultaneously), the increased number of blue, gas-rich neighbours near Type-2 AGN, and the transformation between spiral Type-2 AGN becoming an elliptical Type-1 AGN during the final stages of an interaction. As we can only see what is left after a merger, it is a natural consequence that we would observe fewer galaxies around
the Type-1 AGN. Extending their scenario with a consideration of the morphology of the AGN hosts, we speculate that the blowout of gas would also have to affect some of the neighbours that do not complete the merger, and could provide a merger-driven AGN-feedback mechanism transforming blue nearby galaxies to red ones. In such a case, the ratio between the increase of blue galaxies around Type-2 AGN at close separations $d$ $<$ 50 kpc must be the same or higher than the increase of red galaxies (since some merge) around Type-1 AGN. Also, the bimodality of the colour distribution of galaxies in the Universe should in such a case have a clear correlation with the Type-1/Type-2 ratio evolution over redshift. For the closest bin in the volume-limited cuts of the photometric pair samples, we obtain 783 red and 83 blue galaxies around Type-1 AGN and 1283 blue and 247 red galaxies around Type-2 AGN. Comparing these numbers to those ranging over the complete distance scale up to 350 kpc in Table \ref{SampleSFRAGN}, we see no significant increase in the number of red galaxies around Type-1 AGN, while we see a huge increase of blue neighbours around Type-2 AGN, something which supports our hypothesis.

A problem with the Krongold-Koulouridis scenario is that on average our Type-1 AGN lie at higher redshifts than the Type-2 AGN, both in the parent samples and the neighbour samples. This is consistent with the large number of quasars and deficiency of narrow-line AGN at high redshifts. However, the lack of elliptical Type-2 AGN both in the parent sample and in the pair sample suggests that a Type-2 AGN might be a very short-lived phenomenon, which explains why we have a much smaller probability to observe them in an environment where mergers are more common than now, e.g. in the high-redshift Universe. The spiral Type-1 AGN hosts and elliptical Type-2 AGN hosts would therefore present a form of transition objects between the two types. From this aspect, it would be particularly interesting to study these transition objects in terms of intrinsic properties such as star-formation rate, stellar ages, metallicity and eventual radio loudness. If the time-evolution scenario proves to be correct, we estimate that the ratio between Type-1 spirals and Type-2 ellipticals should remain constant over redshift. Also, we would in such a case expect to see that $>$$>$ 50\% of the Type-2 ellipticals reveal a broad-line region with spectrapolarimetric observations.

The third and the final problem, remains the observation of AGN switching between broad-line and narrow-line mode, which can not be explained successfully by the Koulouridis-Krongold model at present.

\subsection{Final remarks}

We have demonstrated in the paper that the influence of active galaxies on close neighbours largely depends on the nature of the AGN--whether it is a broad-line or a narrow-line AGN. This is not in agreement with the model of AGN unification that has been assumed in many statistical studies of active galaxies, including those with contradicting results of the AGN-starburst or AGN-postburst connection. We suggest as an alternative to separate the two major types of active galaxies when working with large data-sets for easier comparison between various works, and to avoid averaging out possible physical differences in how they interact with close objects.

\section{Acknowledgements}

B. Villarroel wishes to thank Jesus Gonzales (UNAM), Ernst van Groningen (Uppsala Universitet), Allen Joel Anderson (Santa Barbara) and Pianist for great discussions on the work. This work was funded and supported by the Center of Interdisciplinary Mathematics (Uppsala Universitet), the Crafoord foundation and the Swedish Royal Academy of Sciences.


\end{document}